\documentclass[
superscriptaddress,
aps,
reprint,
amssymb, amsmath,
prl,
]{revtex4-1}

\usepackage{newtxtext,newtxmath}
\usepackage[T1]{fontenc}
\usepackage[pdftex]{graphicx}
\usepackage{docs}
\usepackage{bm}
\usepackage{natbib}
\setcitestyle{numbers}
\usepackage{afterpage}
\usepackage{textcomp}
\usepackage{braket}

\usepackage[colorlinks=true,
]{hyperref}%

\newcommand{\myscale}{1.25}

\begin{document}

\title{Complex wavefront engineering with disorder-engineered metasurfaces}

\affiliation{Department of Electrical Engineering, California Institute of Technology, 1200 E. California Blvd., Pasadena, California 91125, USA}
\affiliation{T. J. Watson Laboratory of Applied Physics, California Institute of Technology, 1200 E. California Blvd., Pasadena, California 91125, USA}

\author{Mooseok Jang}
\thanks{These authors contributed equally to this work.}
\affiliation{Department of Electrical Engineering, California Institute of Technology, 1200 E. California Blvd., Pasadena, California 91125, USA}
\affiliation{Present address: Department of Physics, Korea University, 145 Anam-ro, Seongbuk-gu, Seoul 02841, South Korea.}

\author{Yu Horie}
\thanks{These authors contributed equally to this work.}
\affiliation{T. J. Watson Laboratory of Applied Physics, California Institute of Technology, 1200 E. California Blvd., Pasadena, California 91125, USA}

\author{Atsushi Shibukawa}
\thanks{These authors contributed equally to this work.}
\affiliation{Department of Electrical Engineering, California Institute of Technology, 1200 E. California Blvd., Pasadena, California 91125, USA}

\author{Joshua Brake}
\affiliation{Department of Electrical Engineering, California Institute of Technology, 1200 E. California Blvd., Pasadena, California 91125, USA}

\author{Yan Liu}
\affiliation{Department of Electrical Engineering, California Institute of Technology, 1200 E. California Blvd., Pasadena, California 91125, USA}

\author{Seyedeh Mahsa Kamali}
\affiliation{T. J. Watson Laboratory of Applied Physics, California Institute of Technology, 1200 E. California Blvd., Pasadena, California 91125, USA}

\author{Amir Arbabi}
\affiliation{T. J. Watson Laboratory of Applied Physics, California Institute of Technology, 1200 E. California Blvd., Pasadena, California 91125, USA}
\affiliation{Present address: Department of Electrical and Computer Engineering, University of Massachusetts, 151 Holdsworth Way, Amherst, Massachusetts 01003, USA.}

\author{Haowen Ruan}
\affiliation{Department of Electrical Engineering, California Institute of Technology, 1200 E. California Blvd., Pasadena, California 91125, USA}

\author{Andrei Faraon}
\email{faraon@caltech.edu}
\affiliation{T. J. Watson Laboratory of Applied Physics, California Institute of Technology, 1200 E. California Blvd., Pasadena, California 91125, USA}

\author{Changhuei Yang}
\email{chyang@caltech.edu}
\affiliation{Department of Electrical Engineering, California Institute of Technology, 1200 E. California Blvd., Pasadena, California 91125, USA}

\begin{abstract}

  Recently, complex wavefront engineering with disordered media has demonstrated optical manipulation capabilities beyond those of conventional optics. These capabilities include extended volume, aberration-free focusing and subwavelength focusing via evanescent mode coupling. However, translating these capabilities to useful applications has remained challenging as the input-output characteristics of the disordered media ($P$~variables) need to be exhaustively determined via $\mathcal{O}(P)$ measurements. Here, we propose a paradigm shift where the disorder is specifically designed so that its exact characteristics are known, resulting in an \textit{a~priori} determined transmission matrix that can be utilized with only a few alignment steps. We implement this concept with a disorder-engineered metasurface, which exhibits additional unique features for complex wavefront engineering such as an unprecedented optical memory effect range, excellent stability, and a tailorable angular scattering profile.

\end{abstract}

\maketitle

\section{Introduction}

Complex wavefront engineering can be best described as a class of methods that allow control of a very large number of optical degrees of freedom, ranging up to hundreds of thousands \cite{Mosk2012}. This sets it apart from the regime of wavefront manipulation in adaptive optics where the corrections are typically performed for aberrations modeled by a relatively small number of Zernike orders \cite{Tyson2010}. As a class of technologies, complex wavefront engineering is particularly well suited for applications involving disordered media. These applications can be broadly divided into two categories. In the first category, wavefront engineering works to overcome intrinsic limitations of the disordered media. Biological tissue is one such example where scattering is a problem, with complex wavefront engineering emerging as a solution to produce a shaped light beam that counteracts multiple scattering and enables imaging and focusing deep inside the tissue \cite{Horstmeyer2015}.

In the second category, disordered media are intentionally introduced in conjunction with wavefront engineering to unlock an optical space with spatial extent~($x$) and frequency content~($\nu$) that is inaccessible using conventional optics \cite{Vellekoop2010,Vellekoop2010a,VanPutten2011,Park2013,Ryu2016,Boniface2016,Yu2017}. One of the first demonstrations of this ability was reported by Vellekoop \textit{et al.}~\cite{Vellekoop2010}, showing that the presence of a disordered medium (e.g.~a scattering white paint layer) between a source and a desired focal plane can actually help render a sharper focus. In related efforts, researchers have also shown that complex wavefront engineering can make use of disordered media to couple propagating and evanescent modes, in turn enabling near-field focusing \cite{VanPutten2011,Park2013}. Recently, there have been more extensive demonstrations combining disordered media with complex wavefront engineering to increase the flexibility of the optical system to, for example, significantly extend the volumetric range in which aberration-free focusing can be achieved \cite{Ryu2016,Boniface2016,Yu2017}.

Unfortunately, this class of methods is stymied by one overriding challenge -- the optical input-output response of the disordered medium needs to be exhaustively characterized before use \cite{Yu2017,Choi2011,Popoff2011,Kim2015,Popoff2010}. Fundamentally, characterizing $P$ input-output relationships of a disordered medium requires $\mathcal{O}(P)$ measurements. For most practical applications, $P$ greater than $10^{12}$ is highly desired to enable high fidelity access to the expanded optical space enabled by the disordered media with wavefront engineering. Unfortunately, the time-consuming nature of the measurements and the intrinsic instability of the vast majority of disordered media have limited the ability to achieve high values of $P$. To date, the best $P$ quantification that has been achieved is $\sim 10^8$ with a measurement time of 40\,seconds \cite{Choi2011}.

In this paper, we report the use of a disorder-engineered metasurface (we call this a disordered metasurface for brevity) in place of a conventional disordered medium. The disordered metasurface, which is composed of a 2D array of nano-scatterers that can be freely designed and fabricated, provides the optical `randomness' of conventional disordered media, but in a way that is fully known \textit{a~priori}. Through this approach, we reduce the system characterization to a simple alignment problem. In addition to eliminating the need for extensive characterization measurements, the disordered metasurface platform exhibits a wide optical memory effect range, excellent stability, and a tailorable angular scattering profile -- properties that are highly desirable for complex wavefront engineering but that are missing from conventional disordered media. Using this disorder-engineered metasurface platform, we demonstrate full control over $P = 1.1 \times 10^{13}$ input-output relationships after a simple alignment procedure. To demonstrate this new paradigm for controllably exploiting optical `randomness', we have implemented a disordered metasurface assisted focusing and imaging system that is capable of high NA focusing ($\mathrm{NA} \approx 0.5$) to $\sim 2.2 \times 10^{8}$~points in a field of view (FOV) with a diameter of $\sim 8\,\mathrm{mm}$. In comparison, for the same FOV, a conventional optical system such as an objective lens can at most access one or two orders of magnitude fewer points.

\section{Principles}

The relationship between the input and output optical fields traveling through a disordered medium \cite{Popoff2010} can be generally expressed as
\begin{equation}
  E_o(x_o, y_o) = \iint T(x_o, y_o; x_i, y_i) E_i(x_i, y_i) \, \mathrm{d}x_i \mathrm{d}y_i, \label{TM}
\end{equation}
where $E_i$ is the field at the input plane of the medium, $E_o$ is the field at the output plane of the medium, and $T$ is the impulse response (i.e.~Green's function) connecting $E_i$ at a position ($x_i,y_i$) on the input plane with $E_o$ at a position ($x_o,y_o$) on the output plane. In the context of addressable focal spots with disordered medium assisted complex wavefront engineering, Eq.~(\ref{TM}) is discretized such that $E_o$ is a desired focusing optical field, $E_i$ is the linear combination of independent optical modes controlled by the spatial light modulator (SLM), and $T$ is a matrix (i.e.~the transmission matrix) where each element describes the amplitude and phase relationship between a given input mode and output focal spot. In this scenario, $E_i$ has a dimension of $N$, the number of degrees of freedom in the input field (i.e.~the number of SLM pixels), $E_o$ has a dimension of $M$ given by the number of resolvable spots on the projection plane, and $T$ is a matrix which connects the input and output fields with $P$ elements, where $P = M \times N$. We note that the following concepts and results can be generalized to other applications (e.g.~beam steering or optical vortex generation) simply by switching $E_o$ to an appropriate basis set.

One of the unique and most useful aspects of complex wavefront engineering with disordered media is that it allows access to a broader optical space in both spatial extent~($x$) and frequency content~($\nu$) than the input optical field can conventionally access. For example, when an SLM is used alone, the generated optical field $E_i$ contains a limited range of spatial frequencies due to the large pixel pitch of the SLM ($\nu_x$~or~$\nu_y \leq 1/(2d_{\mathrm{SLM}})$ where $d_{\mathrm{SLM}}$ is the pixel pitch; typically $\sim 10\,$\textmu m). As a consequence, the number of resolvable spots $M$ is identical to the number of controllable degrees of freedom $N$. In contrast, when a disordered medium is placed in the optical path, its strongly scattering nature generates an output field $E_o$ with much higher spatial frequencies given by $\sqrt{\nu_x^2 + \nu_y^2} \leq 1/\lambda$, where $\lambda$ is the wavelength of the light. According to the space-bandwidth product formalism \cite{Lohmann1996}, this means that the number of addressable focal spots $M$ within a given modulation area $S$, is maximally improved to
\begin{equation}
  M = S \times \frac{\pi}{\lambda^2}.
\end{equation}
The scheme for focusing with disordered medium assisted complex wavefront engineering can be understood as the process of combining $N$ independent optical modes to constructively interfere at a desired position on the projection plane \cite{Vellekoop2007,Miller2015,Vellekoop2010}. In general, due to the increased spatial frequency range of the output field, the number of addressable spots $M$ is much larger than the number of degrees of freedom in the input, $N$, and therefore the accessible focal points on the output plane are not independent optical modes (see supplementary~\hyperref[S1]{S1}). Instead, each focal spot exists on top of a background which contains the contributions from the unoptimized optical modes in the output field. Here the contrast $\eta$, the ratio between the intensity transmitted into the focal spot and the surrounding background, is dictated by the number of controlled optical modes in the input, $N$ \cite{Vellekoop2007}. In practical situations where, for instance, the addressed spots are used for imaging or photo-switching, the contrast $\eta$ needs to be sufficiently high to ensure the energy leakage does not harmfully compromise the system performance.

To maximize performance, we can see it is desirable to have as many resolvable spots as possible, each with high contrast. This means that both $M$ and $N$, and in turn $P$, should be as high as possible. Practically, there are two ways to measure the elements -- orthogonal input probing and output phase conjugation (see supplementary~\hyperref[S2]{S2}). In each case, an individual measurement corresponds to a single element in the transmission matrix and is accomplished by determining the field relationship between an input mode and a location on the projection plane. Both still necessitate $\mathcal{O}(P)$ measurements which, when $P$ is large, leads to a prohibitively long measurement time. As a point of reference, if the fast transmission matrix characterization method reported in Ref.~\cite{Choi2011} could be extended without complications, it would still require a measurement time of over 40~days to characterize a transmission matrix with $P = 10^{13}$ elements. In comparison, the stability associated with most conventional disordered media can last only several hours \cite{Vellekoop2007,Choi2014,Park2015}.

In contrast, our disorder-engineered metasurface avoids the measurement problem altogether since all elements of the transmission matrix are known \textit{a~priori}. This means that now the procedure to calibrate the system is simplified from the $\mathcal{O}(P)$ measurements needed to determine the transmission matrix to the small number of alignment steps for the disorder-engineered metasurface and the SLM.

A schematic illustration of the technique is presented in Fig.~\ref{fig1} with the omission of a 4-$f$ imaging system optically conjugating the SLM plane to the disordered metasurface. An SLM structures a collimated incident beam into an optimal wavefront which in turn generates a desired complex output wavefront through the disordered metasurface. Since the transmission matrix is known \textit{a~priori}, the process to focus to a desired location is a simple computation. The optimal incident pattern $E_i^{\mathrm{opt}}$ that encodes the information for a target field     $E_o^{\mathrm{target}}$ is calculated using the concept of phase conjugation (see \hyperref[methods]{materials and methods}). This approach enables us to access the maximum possible number of resolvable spots for complex wavefront engineering for a given modulation area $S$ with the added benefit of control over the scattering properties of the metasurface.

\section{Results}
\subsection{The disorder-engineered metasurface}

The disordered metasurface platform demonstrated in this study shares the same design principles as the conventional metasurfaces that have been previously reported to implement planar optical components \cite{Yu2014,Lin2014,Arbabi2014,Backlund2016,Khorasaninejad2016,Enevet2017}: rationally designed subwavelength scatterers or meta-atoms are arranged on a two-dimensional lattice to purposefully shape optical wavefronts with subwavelength resolution (Fig.~\ref{fig2}\textbf{A}). The disordered metasurface, consisting of Silicon Nitride (SiN$_x$) nanoposts sitting on a fused silica substrate, imparts local and space-variant phase delays with high transmission for the designed wavelength of 532~nm. We designed the phase profile $\phi(x,y)$ of the metasurface in such a way that its angular scattering profile is isotropically distributed over the maximal possible spatial bandwidth of $1/\lambda$, and then chose the width of the individual nanoposts according to the look-up table shown in Fig.~\ref{fig2}\textbf{B} (see \hyperref[methods]{materials and methods} for details). The experimentally measured scattering profile confirms the nearly isotropic scattering property of the disordered metasurface, presenting a scattering profile that fully extends to the spatial frequency of $1/\lambda$ as shown in Fig.~\ref{fig2}\textbf{C}. This platform also allows tailoring of the scattering profile, which can be potentially useful in conjunction with angle-selective optical behaviors such as total internal reflection. Figure~\ref{fig2}\textbf{D} presents the measured scattering profiles of disordered metasurfaces designed to have different angular scattering ranges, corresponding to NAs of $0.3, 0.6, \,\mathrm{and}\,0.9$ (see fig.~\ref{sfig1} for 2D angular scattering profiles).

In addition to a highly isotropic scattering profile, the disordered metasurface also exhibits a very large angular (tilt/tilt) correlation range (also known as the optical memory effect \cite{Feng1988}). The correlation is larger than $0.5$ even up to a tilting angle of $30$\,degrees (Fig.~\ref{fig2}\textbf{E}). In comparison, conventional scattering media commonly used for scattering lenses, such as opal glass and several micron-thick Titanium Dioxide (TiO$_2$) white paint layers, exhibit much narrower correlation ranges of less than $1$\,degree (Fig.~\ref{fig2}\textbf{E}) \cite{Schott2015}. Although ground glass diffusers present a relatively wider correlation range of $\sim 5\,$degrees, their limited angular scattering range makes them less attractive for complex wavefront engineering (see fig.~\ref{sfig2} for angular tilt/tilt measurement setup and correlation profiles).

Moreover, the disordered metasurface is extraordinarily stable. We were able to retain the ability to generate a high quality optical focus from the same metasurface without observable efficiency loss over a period of 75~days by making only minor corrections to the system alignment to compensate for mechanical drift (see fig.~\ref{sfig3}).

\subsection{High NA optical focusing over an extended volume}

We experimentally tested our complex wavefront manipulation scheme in the context of disordered medium assisted focusing and imaging. First, we aligned the disordered metasurface to the SLM by displaying a known pattern on the SLM and correcting the shift and tilt of the metasurface to ensure high correlation between the computed and measured output field. Next, to demonstrate the flexibility of this approach, we reconstructed a converging spherical wave (see \hyperref[methods]{materials and methods} for details) for a wide range of lateral and axial focus positions. Figure~\ref{fig3}\textbf{A} presents the simplified schematic for optical focusing (see also \hyperref[methods]{materials and methods} and fig.~\ref{sfig4} for more details). Figure~\ref{fig3}\textbf{B1-B3} shows the 2D intensity profiles for the foci reconstructed along the optical axis at $z'=1.4$, 2.1, and 3.8\,mm, measured at their focal planes. The corresponding NAs are 0.95, 0.9, and 0.75, respectively. The full width at half maximum (FWHM) spot sizes of the reconstructed foci were 280, 330, 370\,nm, which are nearly diffraction-limited as shown in Fig.~\ref{fig3}\textbf{C}. The intensity profiles are highly symmetric, implying that the converging spherical wavefronts were reconstructed with high fidelity through the disordered metasurface. It is also remarkable that this technique can reliably control the high transverse wavevector components corresponding to an NA of $0.95$, while the SLM used alone can control only those transverse wavevectors associated with an NA of $0.033$.

Figure~\ref{fig3}\textbf{B4-B6} shows the 2D intensity profiles at $x'=0$, 1, 4, and 7\,mm on the fixed focal plane of $z'=3.8$\,mm (corresponding to the on axis NA of $0.75$). Because the disordered metasurface based scattering lens is a singlet lens scheme, the spot size along the $x$-axis increased from 370 to 1500\,nm as the focus was shifted (summarized in Fig.~\ref{fig3}\textbf{D}).

The total number of resolvable spots achievable with the disordered metasurface, $M$, was experimentally determined to be $\sim 4.3 \times 10^8$ based on the plot in Fig.~\ref{fig3}\textbf{D}, exceeding the number of controlled degrees of freedom on the SLM ($N \sim 10^5$) by over $3$ orders of magnitude. The NA of $\sim 0.5$ was also maintained in a lateral FOV with a diameter of $\sim 8$\,mm, resulting in $2.2 \times 10^8$ resolvable focal spots. For the sake of comparison, a high-quality objective lens with an NA of $0.5$ typically has $\sim 10^7$ resolvable spots, an order of magnitude smaller than the number of the spots demonstrated with the disordered metasurface.

With our disordered metasurface platform we control a transmission matrix with a number of elements $P$ given by the product of the number of resolvable focal spots on the output plane and the number of controllable modes in the input. The $P$ we achieved with our system was $1.1\times10^{13}$ which allowed us to address $\sim4.3\times10^8$ focus spots with a contrast factor $\eta$ of $\sim2.5\times10^4$. This value of $P$ is $5$ orders of magnitude higher than what has previously been reported \cite{Choi2011}. These findings testify to the paradigm-shifting advantage that this engineered `randomness' approach brings.

We also experimentally confirmed that even with reduced control over the number of input modes, we can still access the same number of resolvable spots on the output plane, albeit with a reduced contrast. By binning pixels on the SLM, we reduced the number of controlled degrees of freedom on the SLM by up to three orders of magnitude, from $\sim 10^5$ to $\sim 10^2$, and verified that the capability of diffraction-limited focusing over a wide FOV is maintained (see fig.~\ref{sfig5}). Although the same number of focal spots can be addressed, the contrast factor $\eta$ is sacrificed when the number of degrees of control is reduced. Using $\sim 10^2$ degrees of freedom in the input, we achieved a contrast factor of $\sim 70$. This validates that the complex wavefront manipulation scheme assisted by the disordered metasurface can greatly improve the number of addressable focal spots for complex wavefront engineering regardless of the number of degrees of freedom in the input.

\subsection{Wide FOV fluorescence imaging}

Finally, we implemented a scanning fluorescence microscope for high-resolution wide FOV fluorescence imaging (see \hyperref[methods]{materials and methods}, fig.~\ref{sfig4}, and fig.~\ref{sfig6} for detailed procedure). Figure~\ref{fig4}A presents the wide FOV low-resolution fluorescence image of immunofluorescence-labeled parasites (\textit{Giardia lamblia cysts}; see \hyperref[methods]{materials and methods} for sample preparation procedures) captured through the $4\times$ objective lens. As shown in the magnified view in Fig.~\ref{fig4}\textbf{B3}, a typical fluorescent image directly captured with a $4\times$ objective lens was significantly blurred, so that the shape and number of parasites was not discernible from the image. Figure~\ref{fig4}, \textbf{B1, C, and D} presents the fluorescence images obtained with our scanning microscope. The scanned images resolve the fine features of parasites both near the center and the boundary of the $5$-mm wide FOV (Fig.~\ref{fig4}\textbf{D}). Our platform provides the capability for high NA focusing ($\mathrm{NA} \approx 0.5$) within a FOV with a diameter of $\sim 8\,$mm, as shown in Fig.~\ref{fig3}. To validate the performance of our imaging system, we compare it to conventional $20\times$ and $4\times$ objectives. The captured images in Fig.~\ref{fig4} demonstrate that we can achieve the resolution of the $20\times$ objective over the FOV of the $4\times$ objective.

\section{Discussion}

Here we have implemented a disorder-engineered medium using a metasurface platform and demonstrated the benefit of using it for complex wavefront engineering. Our study is the first to propose engineering the entire input-output response of an optical disordered medium, presenting a new approach to disordered media in optics. Allowing complete control of the transmission matrix \textit{a~priori}, the disorder-engineered metasurface fundamentally changes the way we can employ disordered media for complex wavefront engineering. Prior to this study, to control $P$ input-output relationships through a disordered medium, $\mathcal{O}(P)$ calibration measurements were required. In contrast, the disorder-engineered metasurface allows for a transmission matrix with $P$ elements to be fully employed with only a simple alignment procedure.

Although we only demonstrate the reconstruction of spherical wavefronts in this study, our method is generally applicable to produce arbitrary wavefronts for applications such as beam steering, vector beam generation, multiple foci, or even random pattern generation (see fig.~\ref{sfig7} for experimental demonstrations). We anticipate that the large gain in the number of addressable optical focal spots (or equivalently angles or patterns) enabled by our method will substantially improve existing optical techniques such as fluorescence imaging, optical stimulation/lithography \cite{Nikolenko2008,Kim2017}, free space coupling among photonic chips/optical networks \cite{Curtis2012,Bruck2016}, and optical encryption/decryption \cite{Pappu2002}.

In the specific application of focal spot scanning, our basic system consisting of two planar components, a metasurface phase mask and a conventional SLM, offers several advantages. The system is highly scalable and versatile, bypassing the limitations and complexities of using conventional objective lenses. The scalability of the metasurface can be especially useful in achieving ultra-long working distances for high NA focusing. The scheme can also be implemented as a vertically integrated optical device together with electronics \cite{Arbabi2016} (e.g.~a metasurface phase mask on top of a transmissive LCD), providing a compact and robust solution to render a large number of diffraction-limited spots. Furthermore, the concept is applicable over a wide range of the electromagnetic spectrum with the proper choice of low-loss materials for the meta-atoms (e.g.~SiN$_x$ or TiO$_2$ for entire visible \cite{Khorasaninejad2016,Zhan2016} and Si for near infrared wavelengths \cite{Arbabi2014,Fattal2010,Vo2014,Arbabi2016a}), which allows for multiplexing different colors, useful for multicolor fluorescence microscopy and multiphoton excitation microscopy. Finally, the planar design provides a platform to achieve ultra-high NA solid-immersion lenses \cite{Ho2015} or total internal reflection fluorescence (TIRF) excitation \cite{Ambrose1956}, suitable for super-resolution imaging and single-molecule biophysics experiments.

More broadly speaking, we anticipate the ability to customize the design of the disordered metasurface for a particular application will prove highly useful. For example, we can tailor the scattering profile of the disordered metasurface to act as an efficient spatial frequency mixer or to be exploited for novel optical detection strategies \cite{Bertolotti2012,Redding2013,Katz2014}. The disordered metasurface can serve as a collection lens, analogous to the results obtained for light manipulation, providing an enhanced resolving power and extended view field. Additionally, the metasurface platform can be designed independently for orthogonal polarization states, which provides additional avenues for control in complex wavefront engineering \cite{Arbabi2015}. Together, the engineering flexibility provided by these parameters offers unprecedented control over complex patterned illumination, which can directly benefit emerging imaging methods that rely on complex structured illumination \cite{Mudry2012,Li2015}.

To conclude, we explored the use of a disorder-engineered metasurface in complex wavefront engineering, challenging a prevailing view of the `randomness' of disordered media by programmatically designing its `randomness'. The presented technology has the potential to provide a game-changing shift that unlocks the benefits of complex wavefront engineering, opening new avenues for the design of optical systems and enabling new techniques for exploring complex biological systems.

\section*{Materials and Methods}\label{methods}
\subsection*{Design of disordered metasurface}
The disordered metasurface consists of Silicon Nitride (SiN$_x$) nanoposts arranged on a subwavelength square lattice with a periodicity of 350\,nm as shown in Fig.~\ref{fig2}\textbf{A}. The width of each SiN$_x$ nanopost is precisely controlled within a range from 60\,nm to 275\,nm, correspondingly imparting local and space-variant phase delays covering a full range of $2\pi$ with close to unity transmittance for an incident wavefront at the design wavelength of 532\,nm (Fig.~\ref{fig2}\textbf{B}). The widths of the nanoposts corresponding to the grayed regions in Fig.~\ref{fig2}\textbf{B} correspond to high quality factor resonances and are excluded in the design of the disordered metasurface. The phase profile $\phi(x,y)$ of the disordered metasurface is designed to yield an isotropic scattering profile over the desired angular range using the Gerchberg-Saxton (GS) algorithm. The initial phase profile of the far-field is randomly chosen from a uniform distribution between $0$ and $2\pi$ radians. After several iterations, the phase profile converges such that the far-field pattern has isotropic scattering over the target angular ranges. This approach helps to minimize undiffracted light and evenly distribute the input energy over the whole angular range.

\subsection*{Fabrication of disordered metasurface}
A SiN$_x$ thin film of 630\,nm is deposited using plasma enhanced chemical vapor deposition (PECVD) on a fused silica substrate. The metasurface pattern is first defined in ZEP520A positive resist using an electron beam lithography system. After developing the resist, the pattern is transferred onto a 60\,nm-thick aluminum oxide (Al$_2$O$_3$) layer deposited by electron beam evaporation using the lift-off technique. The patterned Al$_2$O$_3$ serves as a hard mask for the dry etching of the 630\,nm-thick SiN$_x$ layer in a mixture of C$_4$F$_8$ and SF$_6$ plasma and is finally removed by a mixture of ammonium hydroxide and hydrogen peroxide at $80^\circ\mathrm{C}$.

\subsection*{Alignment procedure}
The alignment procedure consists of two steps to ensure the proper mapping of the SLM pixels onto the intended coordinates of the disordered metasurface. Cross-shaped markers engraved at the four corners of the metasurface are used to guide rough alignment. Then, the marginal misalignments (e.g.~translation and tip-tilt) and aberrations induced by the 4-$f$ system are corrected. For this purpose, a collimated laser beam (Spectra-Physics, Excelsior 532) is tuned to be incident on the metasurface and the resulting field is measured with phase shifting holography. The residual misalignments and aberrations are then calibrated by comparing the measured complex field with the calculated one and digitally compensating for the misalignment by adding appropriate correction patterns on the SLM.

\subsection*{Procedure for optical focusing}
The optimal incident pattern $E_i^{\mathrm{opt}}$ that encodes the information for a target field $E_o^{\mathrm{target}}$ is calculated based on the concept of phase conjugation using the expression
\begin{align*}
  E_i^{\mathrm{opt}}(x,y) &= \mathcal{L}\left[ \iint T^{\dagger}(x,y;x_o,y_o) E_o^{\mathrm{target}}(x_o,y_o) \, \mathrm{d}x_o\mathrm{d}y_o\right] \\
  &= \mathcal{L}\left[ \mathrm{e}^{-i\phi(x,y)} E_o^{\mathrm{target}}(x,y)\right],
\end{align*}
where $\dagger$ represents the conjugate transpose, and the function $\mathcal{L}$ represents the local spatial average of the ideal phase conjugation field $\iint T^{\dagger}E_o^{\mathrm{target}}\, \mathrm{d}x_o\mathrm{d}y_o$ within the area corresponding to each controlled optical mode on the SLM. To produce a focal spot at $\bm{r}'=(x',y',z')$ in free space, the target field is set to a spherical wavefront:
\begin{equation*}
  E_o^{\mathrm{target}}(x,y) = \exp\left[-i\frac{2\pi}{\lambda}\sqrt{(x-x')^2+(y-y')^2+z'^2}\right],
\end{equation*}
where $z'$ is the focal length. To perform the local spatial average $\mathcal{L}$, a low-pass spatial frequency filter is applied using a fast Fourier transform algorithm so that the SLM can successfully sample the optimal wavefront $E_i^{\mathrm{opt}}$. Finally, the SLM (Pluto, Holoeye) is used for phase-only reconstruction of the complex field $E_i^{\mathrm{opt}}$ within a circular aperture with a 4.3\,mm radius. In order
to measure the focal spot, we use a custom-built microscope setup consisting of $100\times$ objective lens (Olympus, UMPlanFl) with an NA of $0.95$, a tube lens (Nikon, $2\times$, Plan Apo), and a CCD camera (Imaging Source, DFK 23UP031).

\subsection*{Procedure for scanning fluorescence imaging}
The setup of our scanning microscope is shown in fig.~\ref{sfig5}\textbf{C}. For the collection of the scanned fluorescent signal, an imaging system consisting of a $4\times$ objective lens (Olympus, $0.1$NA, Plan N) and tube lens (Thorlabs, AC508-100-A-ML) is used to cover most of the FOV of the scanning microscope. We scan the focal spot created behind the metasurface across the region of interest with a 10\,ms pixel dwell time. A pair of galvanometric mirrors are used to scan $2\times2\,$\textmu m$^2$ patches with a step size of 200\,nm, and the neighboring patches are successively scanned by adding a compensation map on the SLM to correct coma aberrations, instead of exhaustively calculating and refreshing the $E_i^{\mathrm{opt}}$ for every spot. The fluorescent signal is detected by the sCMOS camera (PCO, PCO.edge 5.5) with an exposure time of 7\,ms. The fluorescence signal is extracted from the camera pixels corresponding to the scanned focus position. The imaging time for a $30\times30\,$\textmu m$^2$ area is 5\,min, which can be easily improved by two orders of magnitude using a high-power laser and resonant scanning mirrors.

\subsection*{Immunofluorescence-labeled sample preparation}
As a biological sample, we use microscopic parasites, \textit{Giardia lamblia cysts} (Waterborne, Inc.). Before labeling the Giardia, we first prepare (a) the sample of $10^5$ Giardia in 10\,\textmu L phosphate buffered solution (PBS) in a centrifuge tube, (b) 1\,\textmu g of Giardia lamblia cysts antibody (Invitrogen, MA1-7441) in 100\,\textmu L PBS, and (c) 2\,\textmu g of Goat anti-Mouse IgG (H+L) Secondary Antibody conjugated with Alexa Fluor 532 fluorescent dye (Life Technologies, A-11002) in 100\,\textmu L of PBS. The sample (a) is incubated with a blocking buffer. After the blocking buffer is removed, the sample is again incubated with the Giardia antibody solution (b). The sample is rinsed twice with PBS to remove the Giardia antibody solution. The sample is then incubated with the secondary antibody solution with fluorescent dye (c). Finally, the sample is rinsed twice with PBS to remove the secondary antibody solution. All incubations are carried out for 30\,min at $37^\circ\mathrm{C}$. The sample in $10\,$\textmu L PBS is prepared on a slide with Prolong Gold antifade reagent with DAPI (Life Technologies, P36935) to protect the labeled sample from fading and covered with a coverslip.

\section*{Acknowledgment}
This work is supported by the National Institutes of Health BRAIN Initiative (U01NS090577), and a GIST-Caltech Collaborative Research Proposal (CG2012). Y.H. was supported by a Japan Student Services Organization (JASSO) fellowship. Y.H. and A.A. were also supported by National Science Foundation Grant 1512266 and Samsung Electronics. A.S. was supported by JSPS Overseas Research Fellowships. J.B. was supported by the National Institute of Biomedical Imaging and Bioengineering (F31EB021153) under a Ruth L. Kirschstein National Research Service Award and by the Donna and Benjamin M. Rosen Bioengineering Center. S.M.K. was supported by the DOE ``Light-Material Interactions in Energy Conversion'' Energy Frontier Research Center funded by the US Department of Energy, Office of Science, Office of Basic Energy Sciences under Award no. DE-SC0001293. The device nanofabrication was performed at the Kavli Nanoscience Institute at Caltech.

\section*{Author contributions}
M.J. and Y.H. conceived the initial idea. M.J., Y.H., A.S., J.B., and C.Y. expanded and developed the concept. M.J., Y.H., and A.S. developed theoretical modeling, designed the experiments, and analyzed the experimental data. M.J. and A.S. carried out the optical focusing experiments. Y.H. performed the full-wave simulation and the design on the metasurface. A.S. performed the fluorescence imaging experiment. Y.H., S.M.K., and A.A. fabricated the metasurface phase mask. Y.L. performed the measurements on the optical memory effect, the angular scattering profiles, and the stability. All authors contributed to writing the manuscript. C.Y. and A.F. supervised the project.

%

\onecolumngrid

\renewcommand{\figurename}{\textbf{Figure}}
\renewcommand{\thefigure}{\textbf{\arabic{figure}}}

\afterpage{
\begin{figure*}[p]
\centering
\includegraphics[scale=\myscale]{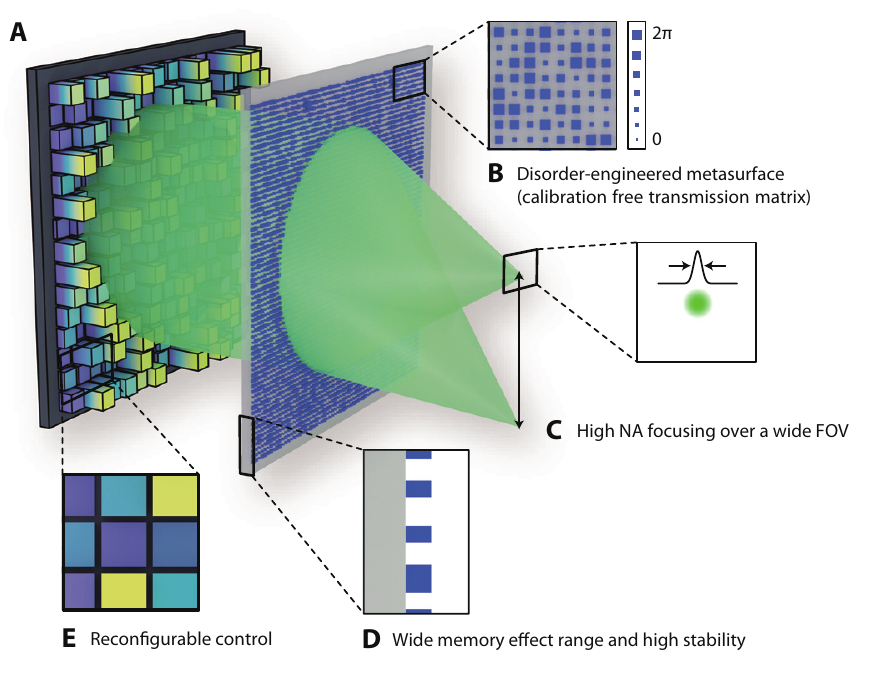}
\caption{
\textbf{Complex wavefront engineering assisted by a disorder-engineered metasurface.}
\textbf{(A)} The system set-up consists of two planar components, an SLM and a disorder-engineered metasurface.
\textbf{(B)} The disorder-engineered metasurface is implemented by varying the size of nanoposts, which correspond to different phase delays $\phi(x,y)$ on the metasurface.
\textbf{(C)} The wide angular scattering range enables high NA focusing over a wide FOV.
\textbf{(D)} The thin, planar nature of the disordered metasurface yields a large memory effect range and also makes the transmission matrix of the metasurface extraordinarily stable.
\textbf{(E)} The SLM enables reconfigurable control of the expanded optical space available through the disordered metasurface.
}
\label{fig1}
\end{figure*}
\clearpage
}

\afterpage{
\begin{figure*}[p]
\centering
\includegraphics[scale=\myscale]{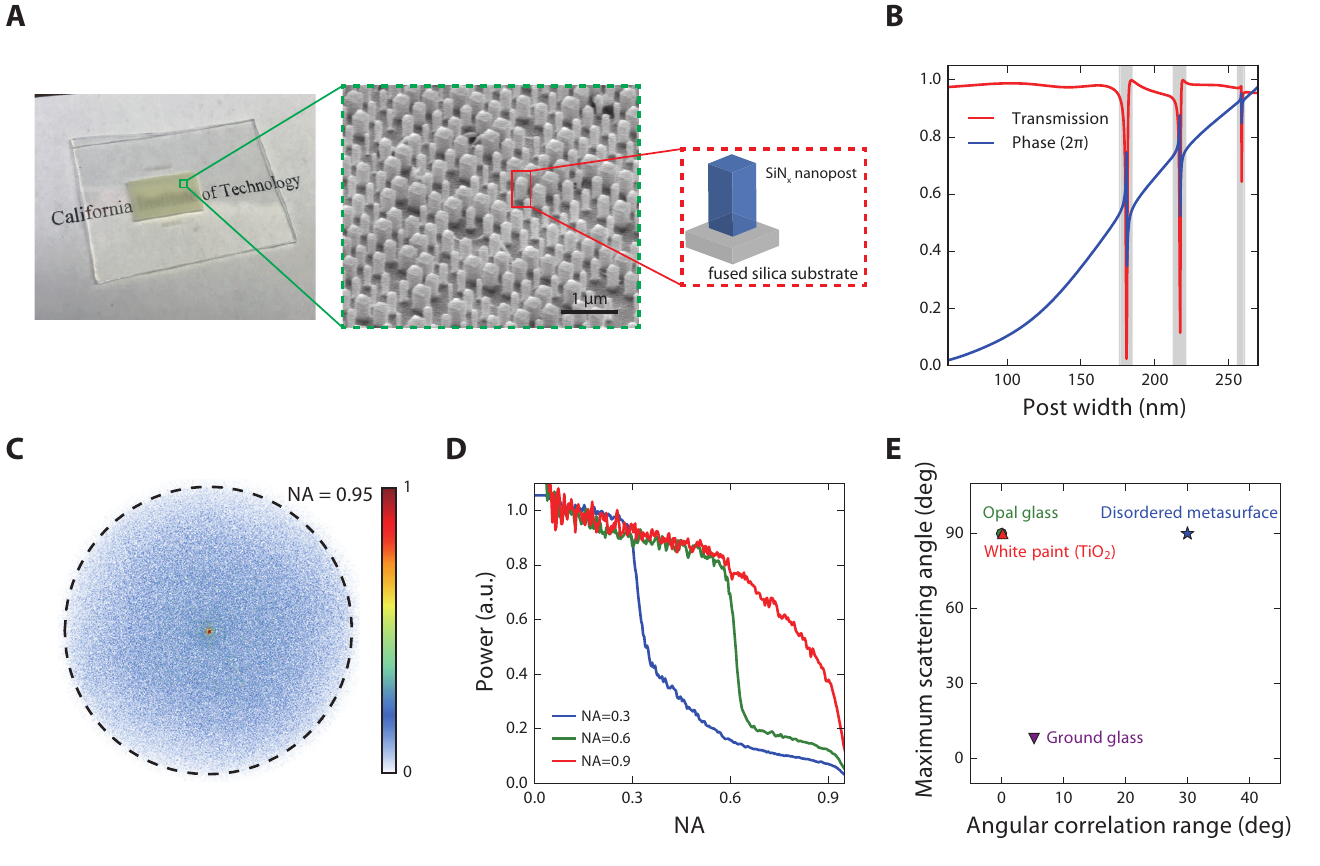}
\caption{
\textbf{Disorder-engineered metasurface.}
\textbf{(A)} Photograph and SEM image of a fabricated disorder-engineered metasurface.
\textbf{(B)} Simulated transmission and phase of the SiN$_x$ nanoposts as a function of their width at a wavelength of 532\,nm. These data are used as a look-up table for the metasurface design.
\textbf{(C)} Measured 2D angular scattering profile of the disordered metasurface, normalized to the strongest scattered field component.
\textbf{(D)} Measured 1D angular scattering profile of the disordered metasurfaces that were specifically designed to scatter the incident light to certain angular ranges ($\mathrm{NA} = 0.3$, $0.6$, $0.9$).
\textbf{(E)} Memory effect range and angular scattering range of the disordered metasurface compared with conventional random media such as white paint, opal glass, and ground glass diffusers.
}
\label{fig2}
\end{figure*}
\clearpage
}

\afterpage{
\begin{figure*}[p]
\centering
\includegraphics[scale=\myscale]{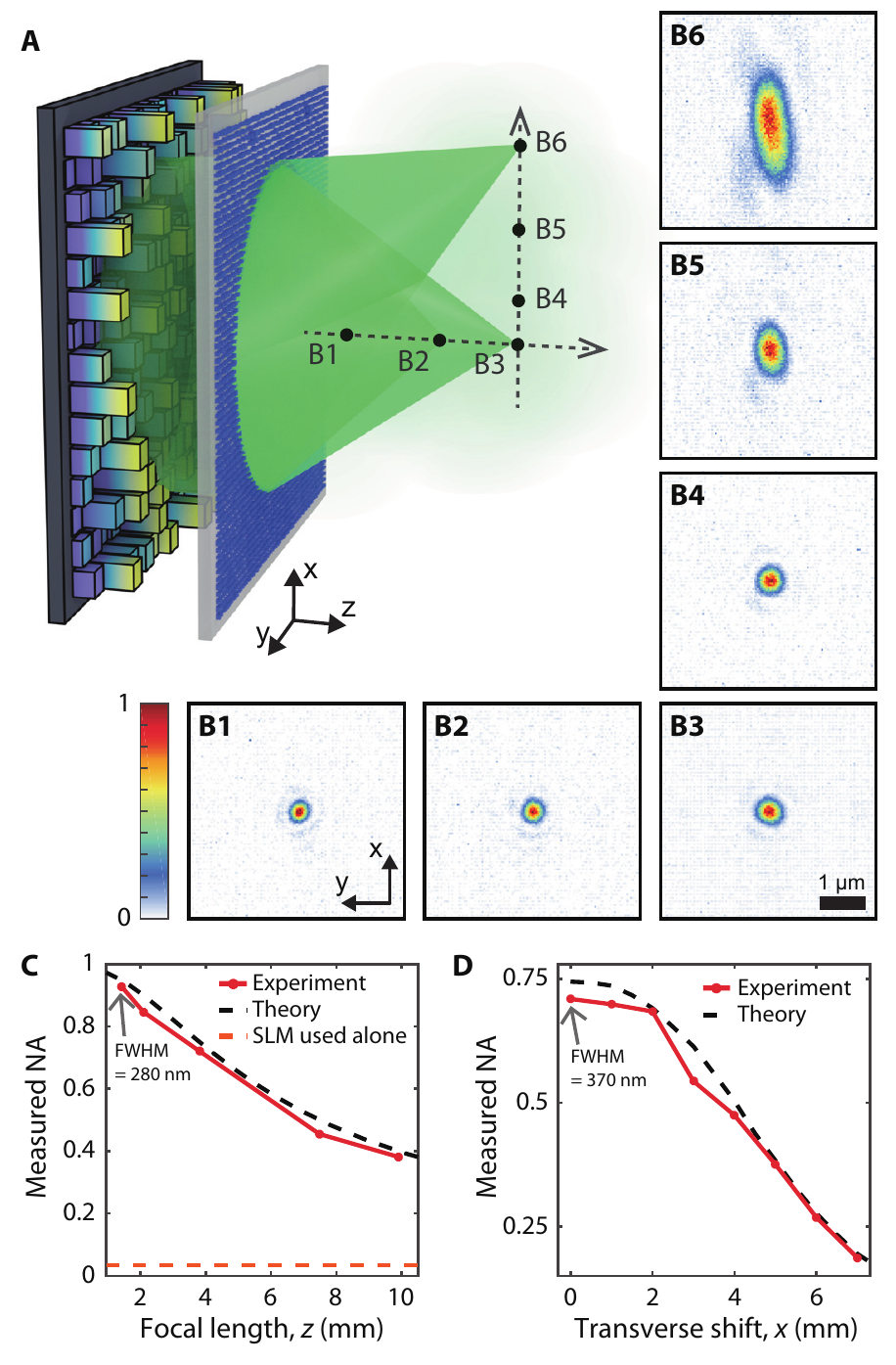}
\caption{
\textbf{Experimental demonstration of diffraction-limited focusing over an extended volume.}
\textbf{(A)} Schematic of optical focusing assisted by the disordered metasurface.
\textbf{(B1-6)} Measured 2D intensity profiles for the foci reconstructed at the positions indicated in \textbf{(A)}. \textbf{B1-B3} are the foci along the optical axis at $z = 1.4$, $2.1$, and $3.8$ mm, respectively, corresponding to NAs of $0.95$, $0.9$, and $0.75$. \textbf{B3-B6} are the foci at $x = 0$, $1$, $4$, and $7$\,mm scanned on the fixed focal plane of $z = 3.8$\,mm. Scale bar: $1\,$\textmu m.
\textbf{(C)} Measured NA (along $x$-axis) of the foci created along the optical axis (red solid line) compared with theoretical values (black dotted line). When the SLM is used alone, the maximum accessible NA is $0.033$ (orange dotted line).
\textbf{(D)} Measured NA (along $x$-axis) of the foci created along $x$ axis at $z = 3.8$\,mm (red solid line) compared with theoretical values (black dotted line). The number of addressable focusing points within the $14$-mm diameter FOV was estimated to be $4.3\times10^8$.
}
\label{fig3}
\end{figure*}
\clearpage
}

\afterpage{
\begin{figure*}[p]
\centering
\includegraphics[scale=\myscale]{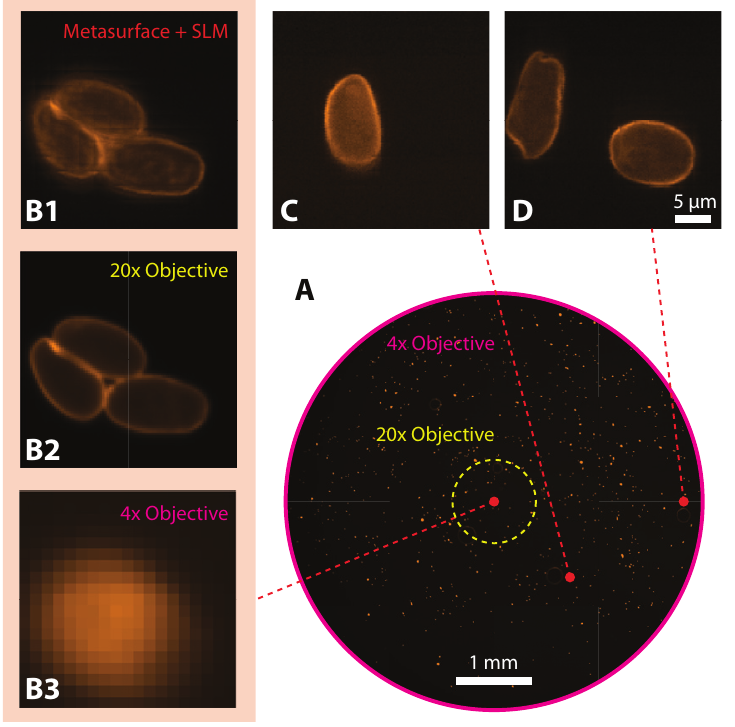}
\caption{
\textbf{Demonstration of disordered metasurface assisted microscope for high resolution wide-FOV fluorescence imaging of \textit{giardia lamblia cysts}.}
\textbf{(A)} Low resolution bright field image captured by a conventional fluorescence microscope with a $4\times$ objective lens ($\mathrm{NA} = 0.1$). Scale bar: $1$\,mm.
\textbf{(B1-3)} Fluorescence images captured at the center of the FOV. \textbf{(B1)} Images obtained with a disordered metasurface lens. \textbf{(B2)} Ground truth fluorescence image captured with a $20\times$ objective lens ($\mathrm{NA} = 0.5$). \textbf{(B3)} Magnified low-resolution fluorescence image captured with the $4\times$ objective.
\textbf{(C, D)} Images obtained with the disorder metasurface-assisted microscope at $(x, y) = (1, 1)$ and $(2.5, 0)$\,mm, respectively. This demonstrates that we can indeed use the system for high resolution and wide FOV imaging.
}
\label{fig4}
\end{figure*}
\clearpage
}

\clearpage

\renewcommand{\thesubsection}{S\arabic{subsection}}
\section*{Supplementary Text}
\subsection{Degrees of freedom in the disordered metasurface assisted wavefront engineering system} \label{S1}

In this supplementary section, we describe the disorder-engineered metasurface and phase-only SLM optical system from the main text in a general mathematical framework. This framework is based on the singular value decomposition (SVD) of the linear operator (e.g.~the transmission matrix, TM), which allows us to rigorously characterize the degrees of freedom of the optical system \cite{Miller2012,Miller2015}. We show that the linear operator connecting the input and output optical modes always has a full rank of $N$ (the number of pixels in the SLM), and thus the degrees of freedom for the output modes is also equal to $N$. However, even though we are limited to $N$ degrees of freedom for the output modes, it is still possible to have a large number of resolvable focal spots within a field of view. Finally, we explain why our system has more degrees of freedom than conventional disordered media.

Any linear optical device can be described by a linear operator $D$ which takes an input function $\ket{\psi_o}$ and generates a linear combination of modes $\ket{\psi_i}$, given as
\begin{equation*}
  \ket{\psi_o} = D \ket{\psi_i},
\end{equation*}
We can always perform the SVD of $D$ which yields
\begin{equation*}
  D = U \Sigma V^\dagger,
\end{equation*}
where $U$ and $V$ are unitary matrices, and $\Sigma$ is a diagonal matrix with complex values that describe the transmission coefficients for independent channels between the input and output modes. By multiplying $U^{\dagger}$ from the left-hand side, we have
\begin{equation*}
  U^\dagger \ket{\psi_o} = \Sigma \left( V^\dagger \ket{\psi_i} \right).
\end{equation*}
The set of modes $U^\dagger \ket{\psi_o}$ and $V^\dagger \ket{\psi_i}$ that correspond with nonzero singular values in $\Sigma$ form the orthogonal sets of basis modes in the output and input spaces.

Next, we consider the case where the linear device operator $D$ represents a general phase mask, the input mode is a wavefront shaped by the SLM, and the output is the field at an arbitrary plane after passing through the phase mask. If the response of the phase mask is insensitive to input angle, the mask can be thought as a device which simply multiplies the input field $\psi_i(x,y)$ by a position-dependent transmission function $T(x,y)$ to obtain the output field $\psi_o(x,y)$ on the device output plane:
\begin{equation*}
  \psi_o(x,y) = T(x,y)\psi_i(x,y).
\end{equation*}
Writing this in matrix form yields
\begin{equation*}
  \ket{\psi_o^p} = D_\mathrm{mask} \ket{\psi_i^p},
\end{equation*}
where we can choose the orthogonal set of input modes as the SLM's pixels. No spatial overlap ensures the orthogonality of the modes. This orthogonality for the $N$ modes holds only if the SLM has a pixel pitch larger than $\lambda/2$. If the pixel pitch is smaller than $\lambda/2$, we cannot count each pixel as an independent mode.

Since the transmission function of the phase mask is local, (i.e.~the phase mask device operation connects an input at a given transverse position on the input plane with an output at the same transverse location on the output plane), the mask operator $D_\mathrm{mask}$ should be diagonal and full-rank in general. This ``local'' effect is not applicable in the case of volumetric scattering media, where an input mode can diffuse inside the media and form a speckle field as an output mode. We will come back to this point later on to compare the two cases. For the corresponding set of output modes $\ket{\psi_o^p}$, the mode orthogonality still holds because the locally transmitted output modes do not spatially overlap right after they are transmitted through the mask.

Describing the optical system of the phase mask and phase-only SLM in this fashion, we
return to the SVD analysis for the system where $D_\mathrm{mask}$ is a diagonal matrix with the elements corresponding to the local transmission coefficients (or, transmission coefficients for the eigenchannels) and $\ket{\psi_i^p}$ and $\ket{\psi_o^p}$ are the pairs of the orthogonal input and output modes respectively. From this SVD analysis, we can see that the device operator (or TM) describing our proposed optical system is always full-rank and we have $N$ degrees of freedom for the output modes as well. This statement is true however one designs the phase mask and however the bases are chosen.

For example, for our disordered metasurface phase mask, we know that plane wave illumination as an input mode can excite all the possible output plane waves nearly isotopically (See Fig.~\ref{fig2}\textbf{C} in the main text). If we describe the system using plane waves as the bases and discretize the angle of the plane waves into $M$ and
$N$ values for the output and input modes, where $M$ is greater than $N$, we can describe the system in the form
\begin{equation*}
  \ket{\psi'_o} = D'_\mathrm{mask} \ket{\psi'_i},
\end{equation*}
where $\ket{\psi'_i}$ and $\ket{\psi'_o}$ are input and output plane wave modes, and $D'_\mathrm{mask}$ is another representation of the device operator $D_\mathrm{mask}$. However, since the description of the system with the operator $D_\mathrm{mask}$ and the input and output sets of orthogonal modes $\ket{\psi^p_i}$ and $\ket{\psi^p_o}$ is a unique and complete characterization of the system, performing the SVD of $D'_\mathrm{mask}$ will result in the same full-rank diagonal matrix $D_\mathrm{mask}$ described above.

So far, we have considered only the linear system describing the field transformation before and after the phase mask. In our experimental scheme, light also propagates from the phase mask to the focal plane. However, free-space propagation can be considered by incorporating the free-space propagation operator, which does not degrade the full-rank operation since it is always full-rank as well.

Now we know that through the metasurface we can control $N$ output modes because we have $N$ degrees of freedom in the input. On the other hand, we also know that we can focus light to a large number of diffraction-limited spots using wavefront engineering (i.e.~choosing the optimum phase for the $N$ input modes in order to form constructive interference peaks at locations of interest). When a disordered medium is used in this way, it is called a ``scattering lens.'' If each resolvable focal spot in the output space is treated as one mode (the total number of which is defined as $M$ according to the space-bandwidth product formalism in the main text), we would seemingly be able to achieve a number of degrees of freedom larger than the rank of our linear system. However, it is not valid to count each resolvable focal spot as an independent mode, because the focal spots created by the scattering lens have correlated, speckle-like backgrounds. Although the number of resolvable focal spots is not equivalent to the number of degrees of freedom, it is an important and useful parameter in many applications. In our focus-scanning scattering lens microscope, since the intensity of an achieved focal spot is significantly higher ($>10^4$) than the background intensity, we can count the number of resolvable focal spots.

It is also worthwhile to analyze the number of degrees of freedom (or eigenchannels) supported by our disordered metasurface phase mask compared to conventional disordered media. For a conventional random medium, multiple scattering processes completely scramble the input modes and generate spatially extended speckle-fields as output modes. In contrast to the mask-based device, the device operator $D_s$ (or TM, with $P = M \times N$ entries) of such a scattering medium is fully populated with complex entries. Similarly, performing the SVD of the TM reveals the number of independent channels for the disordered medium. The TM is generally not full-rank ($\mathrm{rank}(D_s) \leq \min(N, M)$), and it is well-known that the singular value distributions of volumetric disordered media statistically follow the ``quarter-circle law'', experimentally confirmed by Popoff \textit{et al.} \cite{Popoff2010}. Therefore, conventional disordered media deteriorate some degrees of freedom for the output modes, degrading the signal-to-noise ratio (SNR) and the focal contrast $\eta$. This means the advantage of replacing conventional disordered media with a disordered metasurface for complex wavefront engineering is not only that we can operate a scattering lens without characterizing the entire TM of the system, but also that the device operator does not deteriorate the supported degrees of freedom.

\subsection{Conventional measurement of the transmission matrix using $\mathcal{O}(P)$ measurements} \label{S2}

In previous reports, measurements of the transmission matrix have been performed in one of two ways. The first method can be implemented by displaying $N$ orthogonal patterns on the SLM and recording the output field for each pattern \cite{Popoff2010,Choi2011}. This approach can be understood as measuring the transmission matrix one column at a time, where each column corresponds to one SLM pattern, and each element in the column represents the output field contribution at a unique focal point on the projection plane. To focus to a given point on the projection plane, the pattern displayed on the SLM is selected as a linear combination of the SLM patterns such that the output field constructively interferes at the desired focal point. In the context of phase-only modulation, this means that the phase of each field vector, controlled by their respective pixels on the SLM, is aligned so as to maximize the sum over all the field vectors at that location. In order to enable focusing at all $M$ focal spots, the output field for each SLM pattern must be measured at each of the $M$ focal spot locations.

An alternate way to measure the transmission matrix is using optical phase conjugation \cite{Yaqoob2008}. This scheme is typically implemented by creating a calibration light focus from an external lens positioned at the desired focus location and recording the optical field transmitted in the reverse direction through the disordered medium toward the SLM. Then this procedure is repeated by scanning the focus to all $M$ desired focal spots on the output plane. Mathematically, this approach can be interpreted as measuring the transmission matrix one row at a time, where the elements in each row describe the phase and amplitude relationship between a pixel on the SLM and the desired focal point.

While both of these approaches provide a way to characterize the transmission matrix of a disordered medium, they each suffer from practical limitations that prevent them from being practically useful for achieving control over large transmission matrices ($P > 10^{12}$). These stem from the sheer number of measurements and time required to characterize the transmission matrix. The first method is infeasible for large $M$ due to the lack of commercially available camera sensors with the required number of pixels. Thus far, to the best of our knowledge, the largest reported transmission matrix measured using this method contained $P = 10^8$ elements. While the second method is not limited by the availability of the requisite technology, it requires mechanically scanning the focus to each spot. Assuming the relevant measurement technology existed for both cases, with a measurement speed of $10^8$ measurements (i.e.~transmission matrix elements) per second (equivalent to 5\,megapixels at 100\,frames per second), the measurement for all $P = 10^{13}$ elements in our demonstrated transmission matrix would require a measurement time of over 24\,hours. To make matters worse, conventional disordered media used with wavefront engineering such as white paint made of TiO$_2$ or ZnO nanoparticles have a stability of only several hours \cite{Vellekoop2007,Choi2014,Yoon2015}, so the measured transmission matrix would be invalid by the time the measurement was complete.


\renewcommand{\figurename}{\textbf{Figure}}
\renewcommand{\thefigure}{\textbf{S\arabic{figure}}}
\setcounter{figure}{0}

\begin{figure*}[p]
\centering
\includegraphics[scale=\myscale]{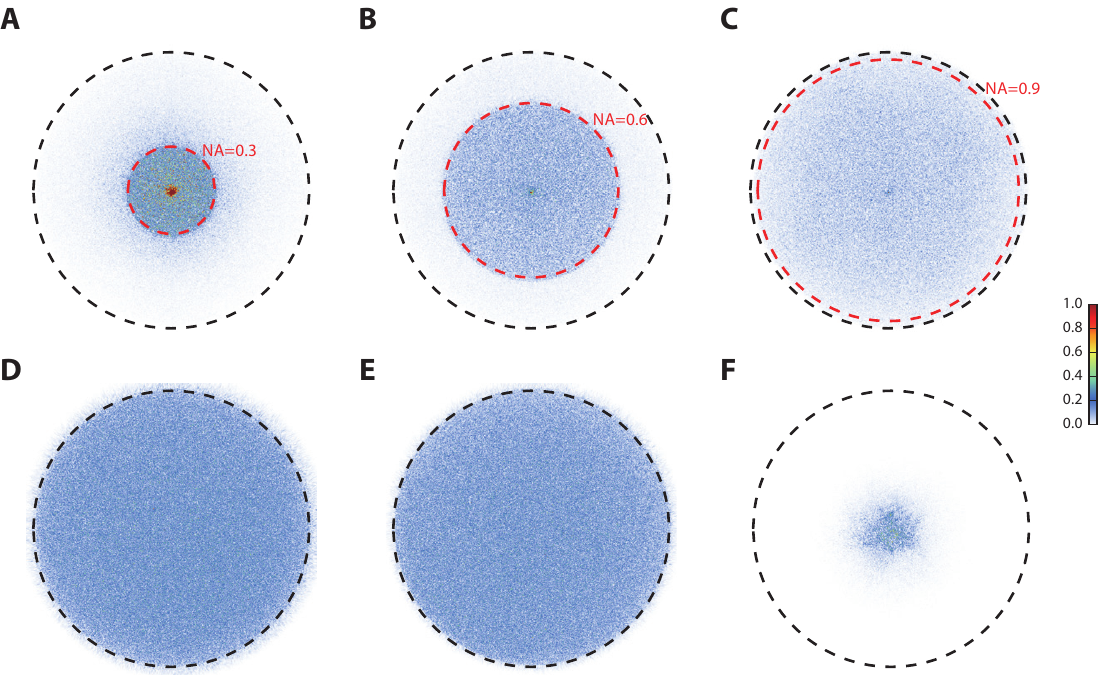}
\caption{
\textbf{Measured angular scattering profiles of disordered metasurfaces as well as those of conventional disordered media.}
A collimated laser beam illuminated the scattering media and a 4-$f$ system imaged the back focal plane of an objective lens ($\mathrm{NA} = 0.95$) to a camera.
\textbf{(A to C)} Angular scattering profiles of disordered metasurfaces with different designs, normalized to strongest scattered field component. The disordered metasurfaces were specifically designed such that they scatter the incident light to certain angular ranges of \textbf{(A)} $\mathrm{NA} = 0.3$, \textbf{(B)} 0.6, \textbf{(C)} 0.9, which are denoted with red dotted lines. See also Fig.~\ref{fig2}\textbf{C} in the main text for the scattering profiles of the disordered metasurface used in the experiment.
\textbf{(D to F)} Angular scattering profiles of conventional scattering media. \textbf{(D)} The 20-\textmu m-thick white paint (made of TiO2 nanoparticles) and \textbf{(E)} opal glass diffuser (10DIFF-VIS, Newport) show isotropic scattering over the wide angular ranges, while \textbf{(F)} the ground glass diffuser (DG10-120, Thorlabs) has a very limited angular range for scattering. The black dotted lines correspond to the cutoff frequencies of the objective lens ($\mathrm{NA} = 0.95$), which is the limit in our measurement set-up.
}
\label{sfig1}
\end{figure*}

\clearpage

\begin{figure*}[t]
\centering
\includegraphics[scale=\myscale]{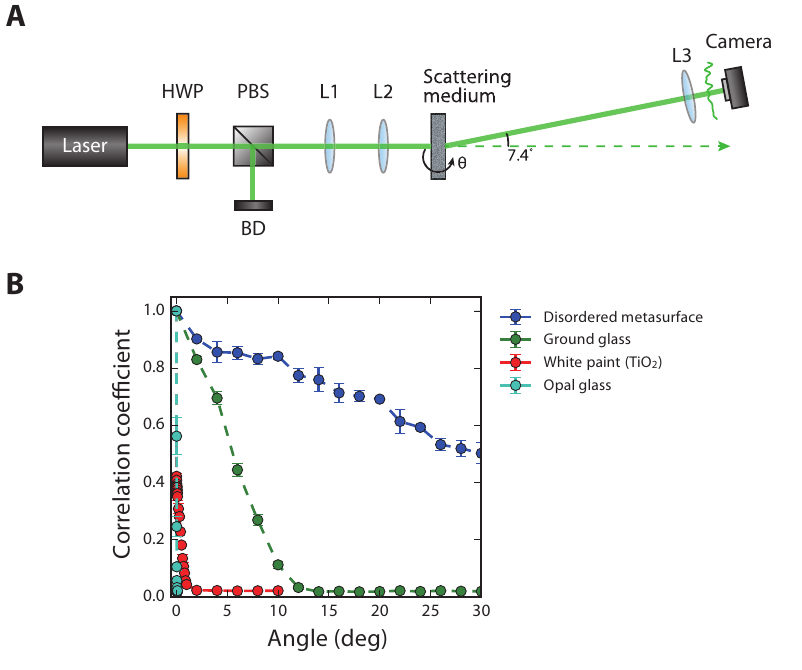}
\caption{
\textbf{Optical memory effect measurement.}
\textbf{(A)} Schematic of the optical set-up to measure the angular correlation range of different scattering media. The output of a long coherence length, 532-nm, continuous-wave laser was attenuated by a variable attenuator composed of a half-wave plate (HWP) and a polarizing beam splitter (PBS) where the unwanted power is sent into a beam dump (BD). After it was expanded to a beam diameter of 8\,mm by lenses L1 and L2, the laser beam illuminated the scattering medium to be tested, and the speckle pattern was detected by a camera. The camera and a camera lens L3 were positioned 7.4\,degrees from the optical axis, to avoid detecting any undiffracted light. The series of speckle patterns were recorded as we rotated the scattering medium, and we computed the correlation coefficient between the first frame and each of the ensuing frames.
\textbf{(B)} The measured memory effect ranges for the disordered metasurface, ground glass (DG10-120, Thorlabs), opal glass (10DIFF-VIS, Newport), and 20-\textmu m-thick white paint (made of TiO$_2$ nanoparticles). See also Fig.~\ref{fig2}\textbf{E} in the main text. Error bars indicate the standard deviation of three measurements.
}
\label{sfig2}
\end{figure*}

\clearpage

\begin{figure*}[t]
\centering
\includegraphics[scale=\myscale]{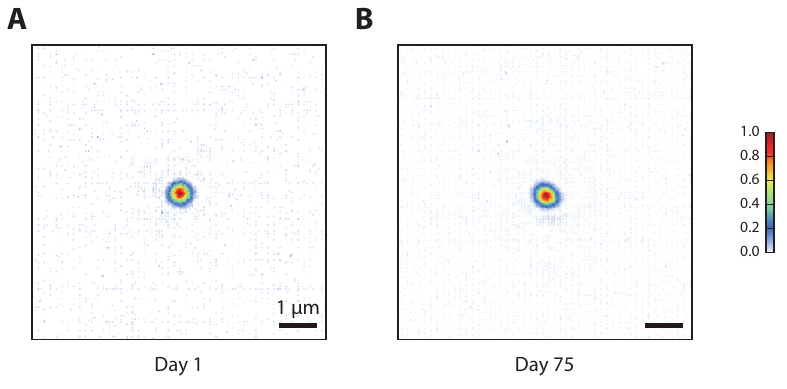}
\caption{
\textbf{Extraordinary stability of a disordered metasurface.}
Over a period of 75 days, a high quality optical focus was obtained from the same metasurface without observable efficiency loss by small system alignments to compensate for mechanical drift.
\textbf{(A)} Reconstructed focus on the 1st day. The measured contrast was 19,800.
\textbf{(B)} Reconstructed focus on the 75th day. The measured contrast was 21,500. Scale bar: 1\,\textmu m.
}
\label{sfig3}
\end{figure*}

\clearpage

\begin{figure*}[t]
\centering
\includegraphics[scale=\myscale]{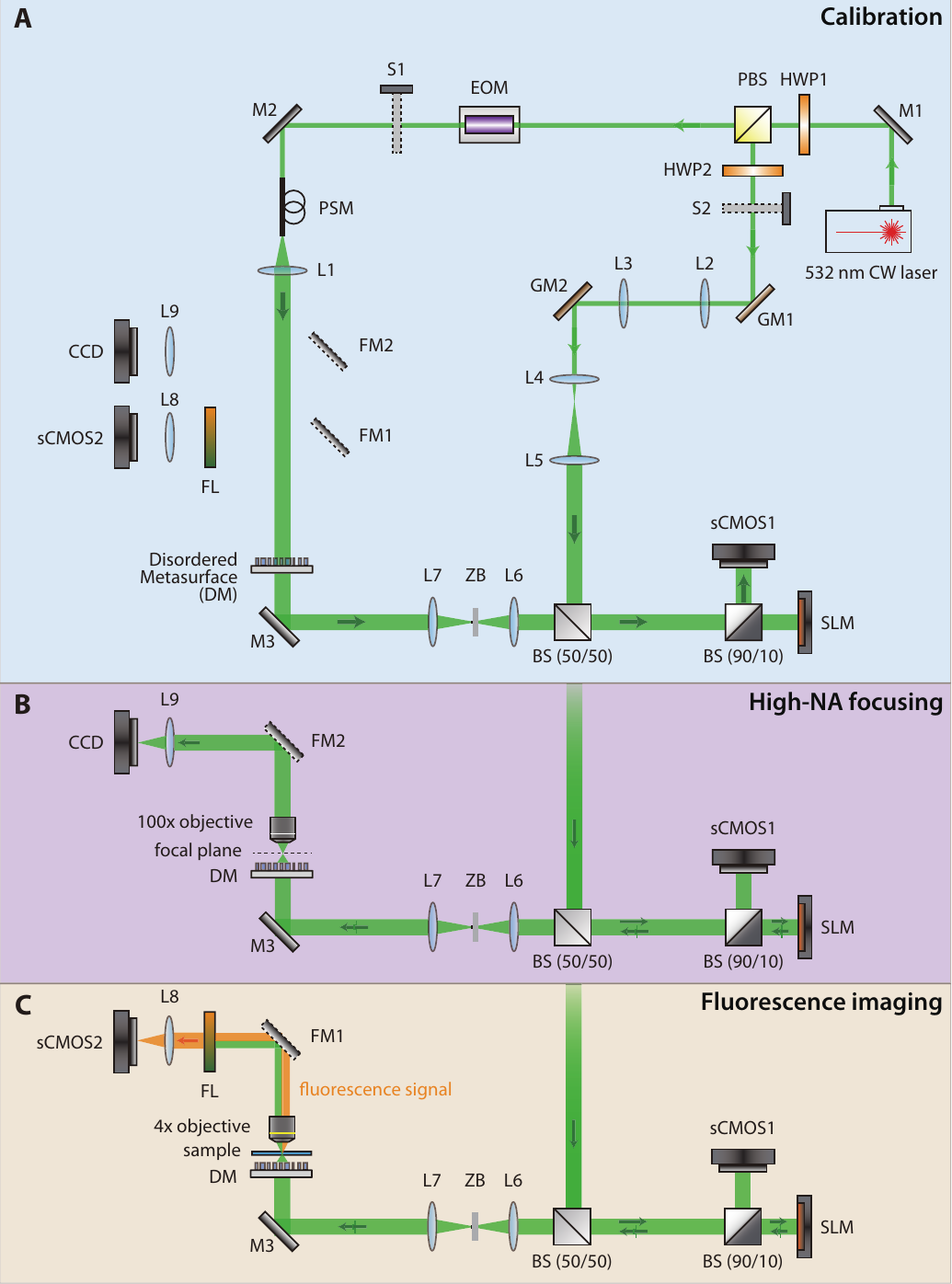}
\caption{
\textbf{Experimental set-up.}
See \hyperref[methods]{materials and methods} for detailed procedures for different experiments.
\textbf{(A)} Phase-shifting holography set-up used for calibrating the alignment for the disordered metasurface and the SLM.
\textbf{(B)} Custom-built microscope set-up used for characterizing high-NA focusing over a wide-FOV.
\textbf{(C)} Focus-scanning fluorescence imaging set-up. M: mirror, L: lens, HWP: half-wave plate, PBS: polarizing beam splitter, S: shutter, EOM: electro-optic modulator, GM: galvanometric mirror, BS: beam splitter, sCMOS: scientific CMOS camera, CCD: CCD camera, SLM: spatial light modulator, ZB: zeroth-order block, DM: disordered metasurface, FM: flip mirror, PSM: polarization-maintaining single-mode fiber, FL: fluorescence filter.
}
\label{sfig4}
\end{figure*}

\clearpage

\begin{figure*}[t]
\centering
\includegraphics[scale=\myscale]{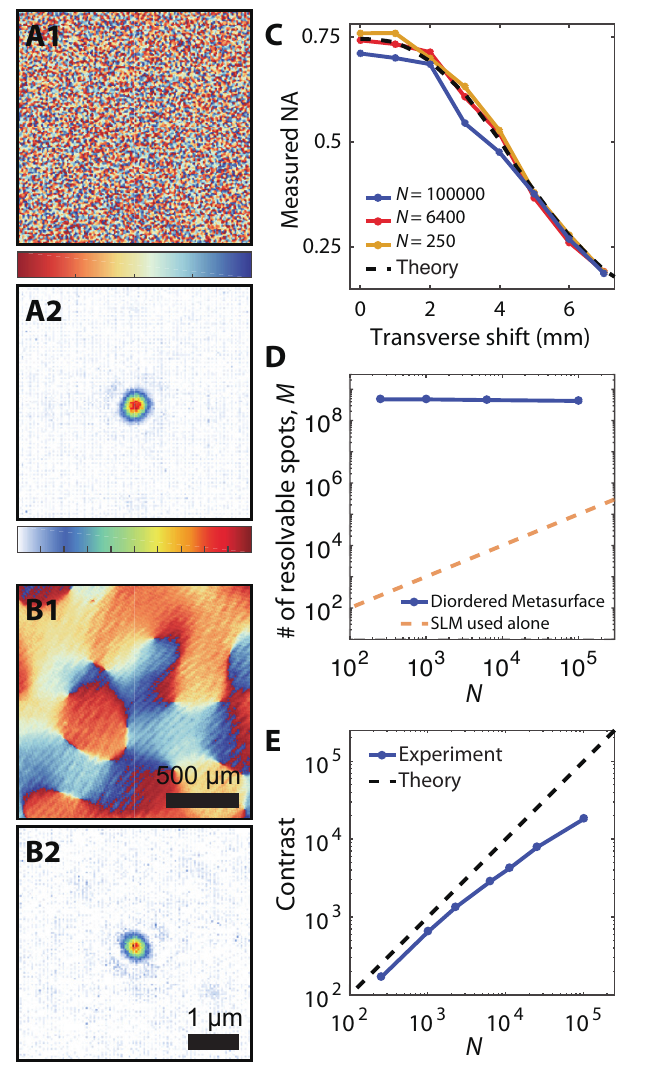}
\caption{
\textbf{Demonstration of ultra-high number of resolvable spots $M$ ($\sim 4.5\times10^8$) even with a handful of physically controlled degrees of freedom ($\sim 2.5\times10^2$) as inputs.}
\textbf{(A1-2, B1-2)} Cropped phase images displayed on the SLM \textbf{(A1, B1)} as well as the corresponding 2D intensity profiles \textbf{(A2, B2)} of the foci reconstructed at $z'=3.8\,$mm on axis ($\mathrm{NA}=0.75$). The controlled number of input optical modes displayed SLM was \textbf{(A1)} $1.0 \times 10^5$ and \textbf{(B1)} $2.5 \times 10^2$, respectively. Scale bars for the phase images and the 2D intensity profiles are 500\,\textmu m and 1\,\textmu m, respectively.
\textbf{(C)} Measured NA of the foci created along $x$-axis. The measured NA shows a good agreement with the theory, regardless of the number of input modes controlled on the SLM.
\textbf{(D)} Measured number of resolvable spots as a function of the number of optical modes controlled on the SLM.
\textbf{(E)} Dependence of contrast factor on the number of optical modes controlled on the SLM.
}
\label{sfig5}
\end{figure*}

\begin{figure*}[p]
\centering
\includegraphics[scale=\myscale]{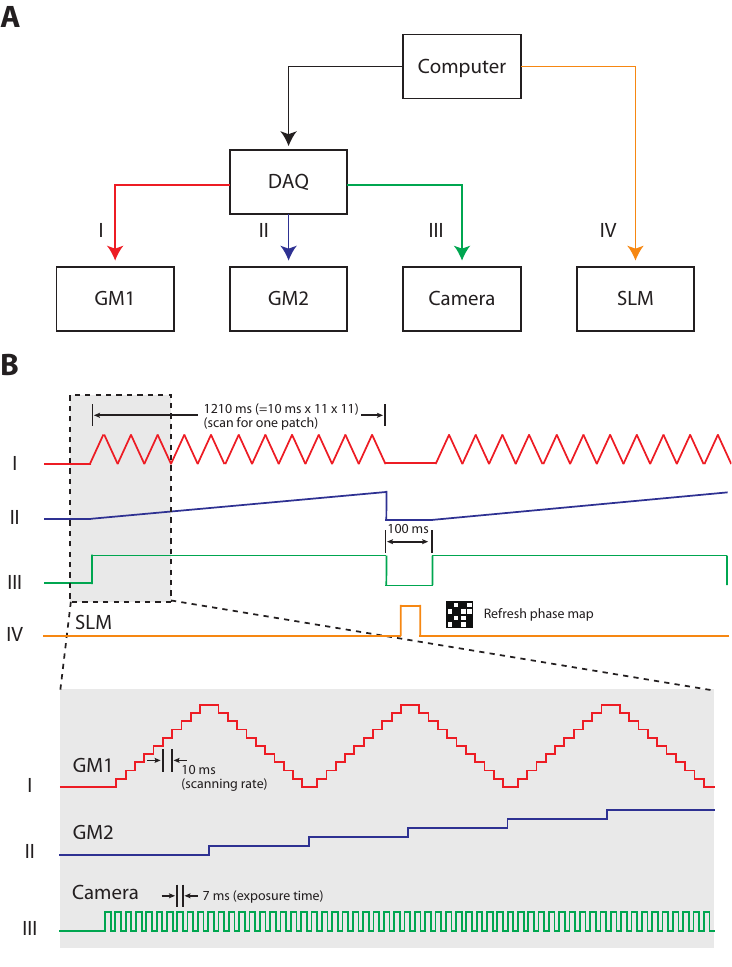}
\caption{
\textbf{Electrical signal flow diagram for scanning fluorescence imaging.}
\textbf{(A)} The system control diagram.
\textbf{(B)} A data acquisition card (DAQ) outputs voltage stepping signals to a pair of galvanometric mirrors (GM1 and GM2) to perform bi-directional raster scanning a pixel dwell time of 10\,ms. At the same time, the DAQ outputs a synchronized trigger signal with a 7\,ms duration to a camera for detecting fluorescent signals. After one patch of $11\times11$ spots has been scanned by the galvanometric mirrors, the galvanometric mirrors return to the original position. During a 100\,ms period, the phase map for correcting coma aberration is updated on a spatial light modulator (SLM). Then, the raster scanning by the galvanometric mirrors is resumed again to constitute another patch.
}
\label{sfig6}
\end{figure*}

\begin{figure*}[p]
\centering
\includegraphics[scale=\myscale]{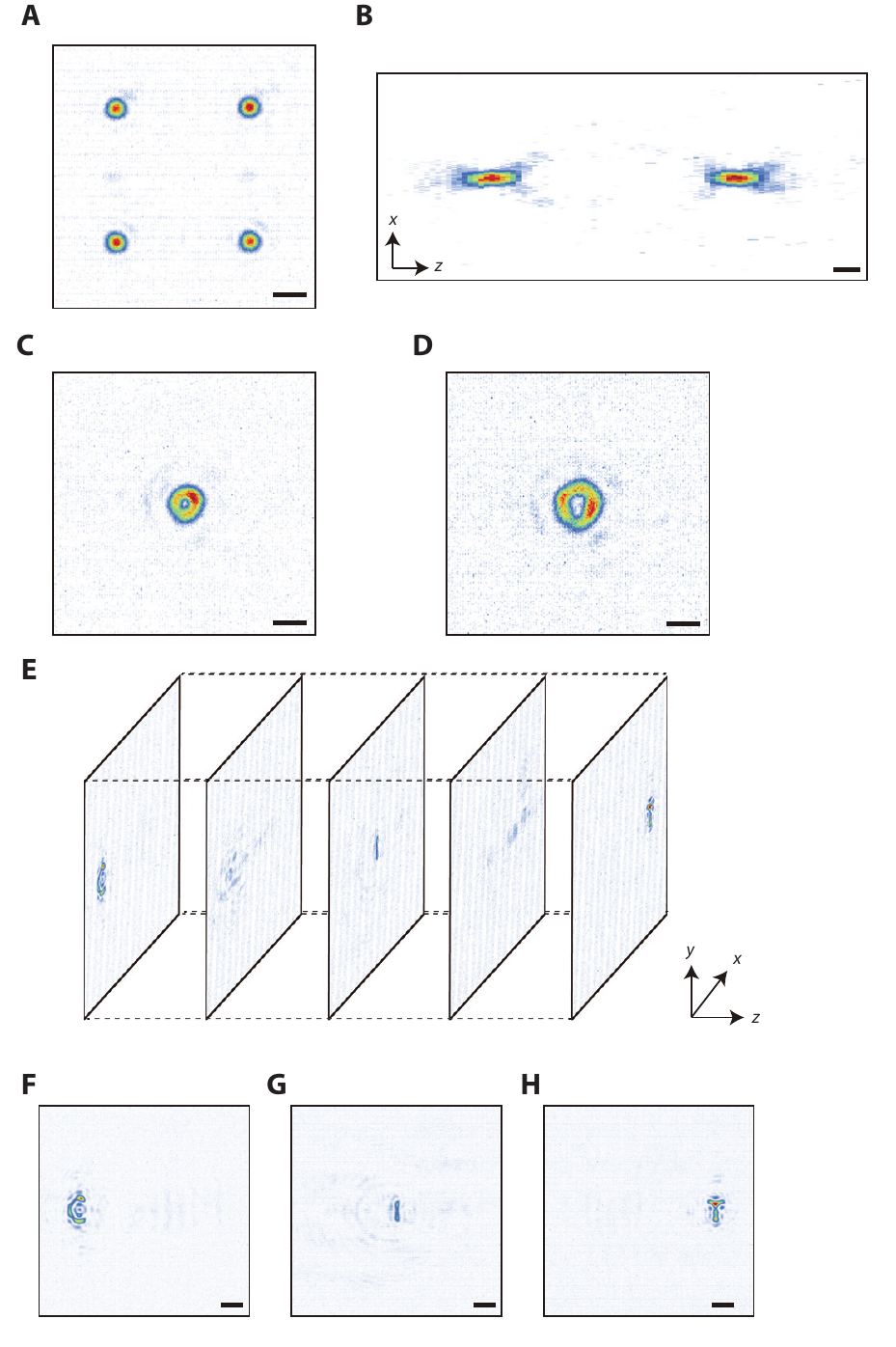}
\caption{
\textbf{Demonstration of arbitrary complex wavefront modulation with a disordered metasurface.}
\textbf{(A, B)} Simultaneous generation of multiple foci. Four foci with 4\,\textmu m distance were reconstructed simultaneously along lateral axis \textbf{(A)}. Two foci with 10\,\textmu m distance were reconstructed simultaneously along optical axis \textbf{(B)}. Scale bar: 1\,\textmu m.
\textbf{(C, D)} Optical vortex focusing with topology charges of $m=1$ \textbf{(C)} and $m=2$ \textbf{(D)}. Scale bar: 1\,\textmu m.
\textbf{(E to H)} 3D display using letters of `C', `I', and `T' placed at \textbf{(F)} $z=-10\,$\textmu m, \textbf{(G)} 0\,\textmu m, and \textbf{(H)} 10\,\textmu m. Scale bar: 2\,\textmu m.
}
\label{sfig7}
\end{figure*}
\end{document}